\documentstyle[graphicx]{mn}

\newcommand{\ls}
 {\mathrel{\hbox{\rlap{\hbox{\lower4pt\hbox{$\sim$}}}\hbox{$<$}}}}
\newcommand{\gs}
 {\mathrel{\hbox{\rlap{\hbox{\lower4pt\hbox{$\sim$}}}\hbox{$>$}}}}

\newcommand{\et}{et al.\ }
\newcommand{\rosat}{{\it ROSAT}}
\newcommand{\logfx}{$\log (f_{\rm X}/f_{\rm V})$}
\newcommand{\logfxfr}{$\log (f_{\rm X}/f_{\rm R})$}
\newcommand{\nh}{$N_{\rm H}$}
\newcommand{\xmm}{{\it XMM}}
\newcommand{\ctof}{$K_{\rm obs}$}
\newcommand{\ctofint}{$K_{\rm int}$}
\newcommand{\ergs}{erg s$^{-1}$}

\newcommand{\einstein}{{\it Einstein}}
\newcommand{\logns}{\mbox{$\log {\it N} - \log {\it S}\:$}}
\newcommand{\asterix}{{\sc asterix}}
\newcommand{\bv}{$B-V$\ }
\newcommand{\br}{$B-R$\ }

\title[X-ray Source Populations in the Region of the Open Clusters
NGC~6633 and IC~4756]
	{X-ray Source Populations in the Region of the Open Clusters
NGC~6633 and IC~4756}
\author[K.R.\ Briggs \et]
	{K.R.\ Briggs$^{1}$, J.P.\ Pye$^1$, R.D.\ Jeffries$^2$,
	E.J.\ Totten$^2$\\
$^1$X-Ray Astronomy Group, Department of Physics and Astronomy, 
	University of Leicester, Leicester LE1 7RH\\
$^2$Department of Physics, Keele University, 
	Keele, Staffordshire, ST5 5BG\\
}
\date{Submitted 03-02-2000}
\date{Resubmitted 08-05-2000}
\date{Resubmitted 18-07-2000}

\pagerange{\pageref{firstpage}--\pageref{lastpage}}
\pubyear{2000}

\begin{document}

\maketitle

\label{firstpage}

\begin{abstract}

Using deep exposures ($\sim 10^5$~s) with the \rosat\ HRI, we have
performed flux-limited surveys for X-ray sources in the vicinity of the
Hyades-age open stellar clusters NGC~6633 and IC~4756, detecting 31 and
13 sources respectively. Our primary aim is to search for so-far
unrecognised cluster members. We propose identifications or
classifications (cluster member, field star, extragalactic field object)
for the X-ray sources, based on published membership lists, and on
X-ray:optical flux ratios and optical colour--magnitude diagrams. Results
of simulating the expected X-ray-emitting source populations are compared
with the \rosat\ measurements and with the expected capabilities of the
\xmm\ mission. The simulations provide a novel method of comparing the
activity levels of NGC~6633 and IC~4756 with that of the Hyades. The
measurements and simulations confirm that cluster members are the major
class of X-ray emitter in these fields at flux levels $ f_{\rm X} \ga \rm
10^{-14}\ erg\ cm^{-2}\ s^{-1} $ (0.1--2.4 keV),
contributing $\sim 40$ per~cent of the total X-ray sources. We find 6
possible new members in NGC~6633, and 4 candidates in IC~4756; all
require further observation to establish membership probability.

\end{abstract}


\section{Introduction}

Open star clusters are central to the study of the evolution of stellar
activity. A stellar cluster provides a coherent sample of stars in terms
of age and chemical-element composition, with a range of masses and
rotational velocities. Comparison between clusters allows the
investigation of stellar properties with age, rotation rate, composition,
and evolutionary environment. The Hyades has long served as one of the
small number of `standard-candle' clusters for much of the detailed study
of the evolution of stellar activity, as representing stars at an age of
$\sim 600$ Myr. However it is not yet established how representative
is any specific cluster of a given age (see e.g.\ the reviews by Randich
1997; Jeffries 1999). In particular, Randich \& Schmitt (1995) reported a
substantial difference in the X-ray luminosity function of cool stars in
Praesepe compared with the Hyades. Apart from the Hyades and Praesepe,
there are only two other well-defined, reasonably compact (angular
diameter $ \la 2^\circ $) open clusters within $\sim 800$ pc, and with
ages in the range $\approx 10^{8.7} $--$10^{8.9} $ yr (c.f.\ Lyng\aa\ 1987),
namely NGC~6633 and IC~4756. Physical data for all four clusters can
be found in Table~\ref{tbl-phys}. Fortuitously, NGC~6633 and IC~4756 lie within a few
degrees of each other in the sky, and at comparable distances from the Sun.

\begin{table*}
\begin{minipage}{150mm}
\caption{Physical properties of NGC~6633 and IC~4756, and comparison
with the Hyades and Praesepe.}
{\footnotesize
\label{tbl-phys}
\begin{tabular}{lcrrrrccr}
Cluster  & $\log$(age) & Dist.\ & Ang.\ Diam.\ & $l$   & $b$     & Metallicity  & $E(B-V)$ & $N_{\rm H}$ \\
Name     &       (yr)      & (pc)   &  (')      &       &         & [Fe/H]       &          &($\rm 10^{20}$$\rm cm^{-2}$)\\
 (1)     & (2)             & (3)    & (4)       & (5)   &  (6)    & (7)            & (8)      & (9)               \\
\\
NGC~6633 & 8.82            & 320    &  27       &  36.1 & $ +8.3$ & 0.0 -- $-0.2$  & 0.17     & 10         \\
IC~4756  & 8.76            & 400    &  52       &  36.4 & $ +5.3$ & +0.04          & 0.19     & 10         \\
\\
Praesepe & 8.82            & 180    &  95       & 205.9 & $+32.5$ & +0.04 -- +0.13 & 0.01     & $\la 1$    \\
Hyades   & 8.82            &  48    & 330       & 180.1 & $-22.3$ & +0.08 -- +0.13 & 0.00     & $\la 0.1$  \\
\end{tabular}
\footnotetext{ 
Data as listed in Lyng\aa\ (1987) except:
(7): Reported range of [Fe/H] from literature; 
(9): Estimated from $E(B-V)$ using $
E(B-V)=1.8 \times 10^{-22}N_{\rm H}$ (Heiles, Kulkarny \& Stark 1981). (8) The
reddening varies across the IC~4756 cluster field from 0.17 to 0.31
(Schmidt~1978) with mean 0.19 (Herzog, Sanders \& Seggewiss 1975) to 0.22
(Schmidt~1978). The reddening across the NGC~6633 cluster field varies
by $\la$ 0.04 (Jeffries 1997).
}
}
\end{minipage}
\end{table*}

The present work focuses on an X-ray observation of NGC~6633 from the
\rosat\ High Resolution Imager (HRI, Tr\"umper \et 1991; David \et 1998).
Our primary aim has been to survey and classify X-ray sources in the
region of NGC~6633, which may be potential unrecognised cluster members,
in a complementary study to the work of Harmer \et (2000) who started
from an optically-selected catalogue of likely members and searched for
X-ray emission from those stars. We have performed a uniform,
flux-limited survey of the X-ray sources in the HRI field containing
NGC~6633, and then, using the X-ray fluxes and two-colour optical
photometry, have attempted to classify each X-ray source as a likely or
possible cluster member, galactic field star or extragalactic object. We
have also compared the observations and proposed empirical
classifications with simulations of the three main expected
X-ray-emitting source populations. For cluster stars, the simulation is
based on a suitably scaled model of the Hyades, thus allowing a direct,
and novel, comparison of activity levels between clusters.

%

\begin{table*}
\begin{minipage}{164mm}
\begin{center}
\caption{\rosat\ observation log.}
\small
\label{tbl-obs}
\begin{tabular}{llllclll}
\rosat\ Observation & Target Name & \multicolumn{2}{c}{Nominal Field Centre} & Exposure & Start Date &  End Date & Original \\
Sequence (ROR)    &             & RA (J2000) & Dec (J2000)                 & (ks)     &            &           & Observer \\       
                  &             & (h~m~s)    & (d~m~s)                     &          &            &           &          \\
\\
rh202056a01 & NGC~6633 & 18 27 31.2 & +06 34 12 & 119 & 13 Sep 1995 & 26 Sep
1995 & J.P. Pye\\
rh202064n00 & IC~4756 & 18 38 31.2 & +05 29 24  & 88  & 16 Sep 1996 & 27 Sep
1996 & K. Singh\\
\end{tabular}
\end{center}
\end{minipage}
\end{table*}

We have performed a similar analysis for IC~4756, albeit with a somewhat
lower exposure in the X-ray image, and with far less optical data. (This
\rosat\ observation has been previously analysed by Randich \et (1998).)

The paper is organized as follows. In Section~2 we describe the X-ray
observations of NGC~6633 and IC~4756, and their analysis. Section~3
discusses the methods adopted for optical identification and
classification, and presents the results. The simulation of clusters (and
field objects) is described in Section~4, while Section~5 extends the
simulations in order to examine the potential of ESA's \xmm\ observatory
(launched in 1999 December) for open-cluster studies. Section~6
summarizes the main points of the paper.

\section{X-Ray Observations and Data Analysis}

Table~\ref{tbl-obs} provides a log of the \rosat\ observations used in this
paper. Table~\ref{tbl-sen} details the sensitivity of these observations.
They are among the deepest HRI exposures within ten degrees of the
Galactic Plane.

\subsection{NGC~6633}


The \rosat\ HRI photon-event list was sorted into an image of approximately
$34 \times 34$ arcmin$^2$ with $1 \times 1$ arcsec$^2$ pixels, using
pulse-height channels 3--8 inclusive (c.f.\ David \et 1998).
This image was searched for point-like sources using the PSS (Point
Source Search) program from the Starlink \asterix\ package (Allan 1992).
PSS uses a maximum-likelihood statistic based on the method of Cash
(1979) to determine the probability that a spatial distribution of counts
arises from a fluctuation in the background, or is indicative of a real
source, and to estimate source parameters (in this case, position and
strength). PSS takes into account the spatial variation (off-axis
broadening) of the HRI point spread function (c.f.\ David \et 1998). 

We modelled the image background initially by smoothing the `raw data'
image using a Gaussian function with full width at half maximum (FWHM)
of 165 arcsec. PSS was run, and detected sources with significance $>$
4.0$\sigma$ were removed from the `raw' image. This source-subtracted
image was Gaussian smoothed in the same way to generate the background
for a second PSS run. Further iterations did not significantly change the
number, significance or counts of the detected sources, so this second
run was the basis for our source detections.

%
\begin{table}
\begin{center}
\caption{Sensitivity of the \rosat\ HRI observations of NGC~6633 and
IC~4756: the fraction of the field (out to 17 arcmin radius from the centre)
covered down to selected
sensitivities. The sensitivities are expressed in terms of HRI on-axis
count rate $C_{\rm X}$ (count/ks), observed (0.1--2.4 keV) flux
$f_{\rm X}$ (in units of $10^{-14}$ erg cm$^{-2}$ s$^{-1}$, using a conversion factor of 3.0$\times
10^{-11}$ erg cm$^{-2}$ s$^{-1}$ per HRI count/s), and intrinsic (0.1--2.4 keV)
luminosity $L_{\rm X}$ for cluster sources (in units of $10^{29}$ erg s$^{-1}$, using a conversion factor of 5.3$\times
10^{-11}$ erg cm$^{-2}$ s$^{-1}$ per HRI count/s and assuming cluster
distances of 320 and 400 pc for NGC~6633 and IC~4756 respectively).}
\small
\label{tbl-sen}
\begin{tabular}{ccccccc}
Fraction of & \multicolumn{3}{c}{NGC~6633} & \multicolumn{3}{c}{IC~4756}\\
field covered & $C_{\rm X}$ & $f_{\rm X}$ & $L_{\rm X}$ & $C_{\rm X}$ & $f_{\rm X}$ & $L_{\rm X}$\\
\\
0.13 & 0.15 & 0.45 & 1.0 & 0.19 & 0.57 & 1.9\\
0.28 & 0.17 & 0.51 & 1.1 & 0.23 & 0.69 & 2.3\\
0.50 & 0.25 & 0.75 & 1.6 & 0.33 & 0.99 & 3.4\\
0.78 & 0.45 & 1.35 & 2.9 & 0.60 & 1.80 & 6.1\\
0.90 & 0.82 & 2.46 & 5.3 & 1.11 & 3.33 &11.2\\
1.00 & 0.93 & 2.79 & 6.0 & 1.25 & 3.75 &12.7\\
\end{tabular}
\end{center}
\end{table}

We generated multiple Poisson random-noise images from our second-run 
background map and established from PSS searches of these images that a significance threshold of 4.5$\sigma$
would yield less than one spurious detection within 17 arcmin of the
field centre.

PSS found 32 significant detections within 17 arcmin of the centre of the
image and returned a source list containing the parameterised position,
99.5 per cent positional error, significance, counts, and 1$\sigma$ count error
for each source. A visual inspection of the image, smoothed with a 10
arcsec FWHM gaussian mask, and of the PSS significance map indicated source
28 to be spurious, part of the extended PSF of the far-offaxis bright source 29. In order to verify the PSS-computed parameters and
errors, sources of known position and counts were added to random-noise
background maps\footnote{ We used the \asterix\ IMSIM program. }, and
PSS was used to find and parameterise the sources in these simulated
fields. We found that the input count was within the computed 1$\sigma$
confidence interval for $\approx$ 80 per cent of the sources, while the
input position was within the PSS-computed symmetric 99.5 per cent confidence
region for only $\approx$ 85 per cent of the sources. An extra 3-arcsec term
added in quadrature to the PSS-computed 99.5 per cent confidence region was
required to correct this. We henceforth refer to this adjusted 99.5 per cent
confidence region as the (positional) {\it error circle}. Effective
on-axis count rates were calculated from the counts, accounting for
detector dead-time, telescope vignetting and spatially-varying quantum
efficiency of the detector. These count rates were converted to observed
fluxes in the 0.1--2.4 keV energy band using a conversion factor (\ctof) of
3.0$\times$10$^{-11}$ erg cm$^{-2}$s$^{-1}$ per HRI count
s$^{-1}$, as acceptable spectral fits have been found for active coronae
using single-temperature Raymond-Smith models with $kT$ in the range
0.5--1.5 keV, and the PIMMS program (provided by NASA's
GSFC/HEASARC) computed conversion factors of 2.8$\times$10$^{-11}$ and
3.1$\times$10$^{-11}$ erg cm$^{-2}$s$^{-1}$
per HRI count s$^{-1}$ for Raymond-Smith models with $kT=0.86$ and
$kT=1.4$ keV respectively (with absorption
column density $N_{\rm H} = 0$). Assuming a power-law spectrum with
photon-index $\gamma$ in the range 1.8--2.2, and/or increasing $N_{\rm
H}$ to 2$\times10^{21}$ $\rm
cm^{-2}$ changes \ctof\ by a factor $\la $1.3.

\begin{figure}
\begin{center}
\rotatebox{-90}{\includegraphics[width=7.8cm]{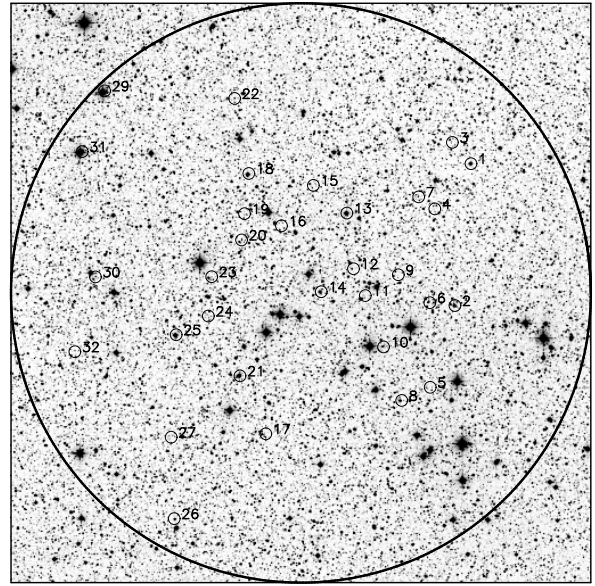}}
\end{center}
\caption{Positions of detected X-ray sources overlaid on the Digitised Sky
Survey image of the NGC~6633 HRI field. The image is 34 arcmin across
and a circle of 17 arcmin radius marks the area considered in our survey.
The image is centred on the approximate cluster centre, at the
coordinates given in Table~\ref{tbl-obs}; North is up and East is to the
left. }
\label{x-opt-image}
\end{figure}

The X-ray characteristics of
the 32 sources detected in the NGC~6633 field are given in Table~\ref{tbl-ngc6633-x}.

Franciosini, Randich \& Pallavicini (2000) have also studied this observation,
but find only 24 sources above a significance threshold of $\approx 3.5
\sigma$. An examination of their Figure~1 indicates that 20 of these are
common to our sourcelist. The lower source count may be due to the larger
pixel sizes (5''$\times$ 5'') and greater range of PHA channels (3--15) they have
taken to form their image. When we produced an image using their
binning and PHA range and ran PSS in the manner described above, we detected only 18 sources above our
4.5$\sigma$ significance threshold.

%
\begin{table*}
\begin{minipage}{90mm}
\caption{X-ray sources in the field of
NGC~6633.}
\label{tbl-ngc6633-x}
\begin{center}
\scriptsize
\begin{tabular}{rcccccc}
Nr & RA (J2000) & Dec (J2000) & Ecirc & Offax & Signif & C$_{\rm X} \pm \sigma_{{\rm C}_{\rm X}}$ \\
(1) & (2) & (3) & (4) & (5) & (6) & (7)\\
\\
 1 &18 26 50.89&+ 6 41 49.0&9.72&12.58 & 4.65 &0.39 $\pm$ 0.10\\
 2 &18 26 54.70&+ 6 33 29.5&6.66& 9.09 & 4.59 &0.20 $\pm$ 0.06\\
 3 &18 26 55.29&+ 6 43 04.3&6.24&12.58 & 5.13 &0.40 $\pm$ 0.10\\
 4 &18 26 59.44&+ 6 39 10.1&4.86& 9.32 & 4.68 &0.20 $\pm$ 0.06\\
 5 &18 27 00.56&+ 6 28 41.7&4.80& 9.39 & 6.33 &0.30 $\pm$ 0.07\\
 6 &18 27 00.64&+ 6 33 39.5&4.50& 7.61 & 7.13 &0.29 $\pm$ 0.06\\
 7 &18 27 03.27&+ 6 39 52.1&4.62& 8.96 & 5.95 &0.22 $\pm$ 0.06\\
 8 &18 27 07.29&+ 6 27 55.3&5.52& 8.64 & 5.02 &0.17 $\pm$ 0.05\\
 9 &18 27 08.07&+ 6 35 18.0&5.88& 5.85 & 5.26 &0.18 $\pm$ 0.05\\
10 &18 27 11.57&+ 6 31 04.1&5.52& 5.80 & 4.88 &0.15 $\pm$ 0.05\\
11 &18 27 15.86&+ 6 34 04.6&5.40& 3.81 & 4.74 &0.15 $\pm$ 0.05\\
12 &18 27 18.65&+ 6 35 39.4&4.92& 3.44 & 4.78 &0.14 $\pm$ 0.04\\
13 &18 27 20.22&+ 6 38 55.2&5.10& 5.45 & 4.71 &0.15 $\pm$ 0.05\\
14 &18 27 26.26&+ 6 34 18.3&3.78& 1.23 & 9.36 &0.36 $\pm$ 0.07\\
15 &18 27 28.13&+ 6 40 32.7&4.14& 6.39 & 9.39 &0.44 $\pm$ 0.08\\
16 &18 27 35.66&+ 6 38 10.4&5.40& 4.13 & 4.56 &0.15 $\pm$ 0.05\\
17 &18 27 39.35&+ 6 25 58.1&4.20& 8.48 & 8.72 &0.38 $\pm$ 0.07\\
18 &18 27 43.36&+ 6 41 14.6&5.88& 7.66 & 5.17 &0.21 $\pm$ 0.06\\
19 &18 27 44.39&+ 6 38 53.0&5.10& 5.72 & 4.75 &0.15 $\pm$ 0.05\\
20 &18 27 45.09&+ 6 37 20.6&4.98& 4.67 & 5.18 &0.17 $\pm$ 0.05\\
21 &18 27 45.44&+ 6 29 22.0&3.30& 5.99 &20.43 &1.14 $\pm$ 0.11\\
22 &18 27 46.72&+ 6 45 39.0&6.96&12.08 & 4.84 &0.41 $\pm$ 0.11\\
23 &18 27 52.14&+ 6 35 11.4&4.98& 5.29 & 4.71 &0.15 $\pm$ 0.05\\
24 &18 27 52.92&+ 6 32 52.7&5.28& 5.55 & 6.15 &0.21 $\pm$ 0.05\\
25 &18 28 00.52&+ 6 31 46.9&5.64& 7.67 & 5.04 &0.18 $\pm$ 0.05\\
26 &18 28 00.96&+ 6 20 59.4&8.94&15.14 & 5.54 &0.76 $\pm$ 0.17\\
27 &18 28 01.70&+ 6 25 46.0&4.68&11.34 & 8.45 &0.45 $\pm$ 0.08\\
28 &18 28 17.33&+ 6 45 34.9&6.24&16.15 & 5.87 &0.84 $\pm$ 0.18\\
29 &18 28 17.43&+ 6 46 05.4&5.16&16.53 &10.45 &1.64 $\pm$ 0.20\\
30 &18 28 19.59&+ 6 35 10.5&5.94&12.06 & 4.76 &0.36 $\pm$ 0.10\\
31 &18 28 22.72&+ 6 42 31.5&5.28&15.26 & 4.90 &0.70 $\pm$ 0.17\\
32 &18 28 24.43&+ 6 30 46.8&4.92&13.66 &10.00 &0.88 $\pm$ 0.12\\
\end{tabular}
\vspace{-0.5cm}
\footnotetext{Key:
(4) Positional error circle radius (arcsec); 
(5) Offaxis angle (arcmin); 
(6) Significance ($\sigma$); 
(7) PSS Count rate and 1$\sigma$ error (HRI counts/ksec)} 
\end{center}
\end{minipage}
\end{table*}

\subsection{IC~4756}

Randich \et (1998) (henceforth R98) found 10 X-ray sources in the HRI image of IC~4756.

Our analysis of the \rosat\ data for the IC~4756 field proceeded in a similar
manner to that for NGC~6633, and we only point out the differences
here. Again, a \ctof\ of
3.0$\times10^{-11}$ erg cm$^{-2}$s$^{-1}$ per HRI count s$^{-1}$ was used to calculate absorbed fluxes.
PSS found 13 sources with significance $> 4.5$ within 17
arcmin of the field centre. Their characteristics are listed in Table~\ref{tbl-ic4756-x}. 9 of these were detected by R98, whose
numbering system appears in column 2 of Table~\ref{tbl-ic4756-x}. The count rates
calculated by PSS average 16 per cent smaller than those quoted by R98, with a standard deviation of 13 per cent. We detected R98's
source 2 at a significance of 4.4$\sigma$, i.e.\ marginally below our
threshold level of 4.5$\sigma$, so it is not included here.

\section{Optical Identification and Classification of Sources}


\subsection{NGC~6633}

Initially a catalogue of stars in the region of NGC~6633 was compiled
from the CDS SIMBAD database and a variety of previous surveys of the
cluster (Kopff 1943; Hiltner, Iriarte \& Johnson 1958; Vasilevskis,
Klemola \&  Preston 1958; Sanders 1973;
Jeffries 1997). This catalogue was very incomplete for $V > 11$,
but was sufficient for an initial assessment, and in particular to check
for any systematic position errors in the X-ray image. The catalogue was
searched for matches within 12 arcsec of the X-ray source positions. Ten
matches were found, of which 7 were known proper-motion members of the
cluster, 1 was a proper-motion non-member and 2 were G-type giants
(Kopff 1943) which had been
assigned low probabilities of membership (Sanders 1973). Two of the
proper-motion members, J25 and S359 (matched with sources 18 and 21
respectively) are recognised spectroscopic binaries (SBs) (Harmer \et
2000). The two-dimensional $rms$ difference between the X-ray-source
and optical-catalogue positions was 3.5
arcsec, and any systematic offset was $<$0.7 arcsec, so no boresight
correction was applied. Fig.~\ref{x-opt-image} shows the relative
positions of the detected X-ray sources, overlaid on a
Digitized Sky Survey (DSS) image of the NGC~6633 HRI field,
highlighting that many of the optically-brightest members of the
cluster are not significant X-ray sources.

%
\begin{table*}
\begin{minipage}{100mm}
\caption{X-ray sources in the field of
IC~4756.}
\label{tbl-ic4756-x}
\begin{center}
\scriptsize
\begin{tabular}{rrcccccc}
Nr & R98 & RA (J2000) & Dec (J2000) & Ecirc & Offax & Signif & $C_{\rm
X}\pm \sigma_{C_{\rm X}}$ \\
(1) & (2) & (3) & (4) & (5) & (6) & (7) & (8) \\
\\
 1 & 1 & 18 38 20.02 & + 5 24 59.0  & 3.93 & 5.22 & 7.85 & 0.37 $\pm$ 0.08 \\
 2 & 3 & 18 38 31.89 & + 5 31 51.8  & 4.16 & 2.47 & 7.20 & 0.34 $\pm$ 0.08 \\
 3 & 4 & 18 38 35.63 & + 5 34 52.4  & 3.79 & 5.58 &10.56 & 0.62 $\pm$ 0.10 \\
 4 & 5 & 18 38 36.94 & + 5 35 45.2  & 5.07 & 6.51 & 5.69 & 0.28 $\pm$ 0.07 \\
 5 & 6 & 18 38 37.81 & + 5 33 59.5  & 4.03 & 4.88 & 7.79 & 0.39 $\pm$ 0.08 \\
 6 &   & 18 38 40.91 & + 5 29 47.5  & 4.93 & 2.45 & 4.59 & 0.18 $\pm$ 0.06 \\
 7 &   & 18 38 47.73 & + 5 41 00.6  & 5.73 &12.32 & 4.73 & 0.49 $\pm$ 0.13 \\
 8 & 7 & 18 38 50.94 & + 5 31 54.8  & 3.34 & 5.52 &14.94 & 0.90 $\pm$ 0.11 \\
 9 & 8 & 18 39 04.67 & + 5 19 35.0  & 5.30 &12.88 & 8.90 & 1.01 $\pm$ 0.16 \\
10 & 9 & 18 39 05.55 & + 5 34 59.8  & 4.60 &10.22 & 8.28 & 0.65 $\pm$ 0.12 \\
11 &10 & 18 39 05.57 & + 5 34 24.1  & 3.80 & 9.91 &13.85 & 1.24 $\pm$ 0.15 \\
12 &   & 18 39 16.06 & + 5 39 29.5  & 5.03 &15.05 & 4.84 & 1.00 $\pm$ 0.25 \\
13 &   & 18 39 21.69 & + 5 22 48.1  & 6.70 &14.19 & 4.76 & 0.48 $\pm$ 0.13 \\
\end{tabular}
\footnotetext{Key:
(2) Source number in Randich \et (1998);
(5) Positional error circle radius(arcsec); 
(6) Offaxis angle(arcmin); 
(7) Significance ($\sigma$); 
(8) PSS count rate and 1$\sigma$ error (HRI counts/ksec)}
\end{center}
\end{minipage}
\end{table*}

Totten, Jeffries \& Hambly (2000) (henceforth TJH) have recently
performed a photometric and spectroscopic study of NGC~6633, complete
down to $V \sim 19$ and extending down to $V \approx 21$. A search of
the TJH catalogue for objects within the error circles of our X-ray
sources found optical candidates for 22 sources\footnote{The giant
we have identified with source 29 falls just outside its error circle,
but the probability of such a bright object appearing so close to an
X-ray source by chance is so small, we consider this to be the correct
identification.}. Photometry and positional information for these potential counterparts
are listed in columns 2--6 of Table~\ref{tbl-ngc6633-opt}. For 8 X-ray sources
there was more than one optical object in the error circle. We would
expect a total of $\sim 16$ catalogued objects to fall into the error
circles simply by chance (the majority of these fainter than
18th magnitude), suggesting several further mis-associations in addition
to the eight obvious (multi-object) cases. The TJH catalogue covers 83 per cent of
the entire field we are using and is incomplete for $V
< 10 $. Photometric and positional data for these bright stars (the two giants)
were taken from the Tycho ACT catalogue (Urban, Corbin \& Wycoff
1997), and are
given in columns 2--6 of Table~\ref{tbl-ngc6633-opt}. Complete
coverage of the field was found by searching the USNO-A2.0 catalogue
(Monet \et 1998), which extends down to $R \sim 20$, but is complete only
to $R \sim 17$: this yielded 21 candidate identifications for 20 sources,
19 of these the same as those found from the TJH/ACT
catalogues. Positional and photometric information are listed in
columns 2--3 and 8--10 of Table~\ref{tbl-ngc6633-opt}. We
would expect nine objects to fall into the error circles simply by
chance.

In all, 26 of the 31 X-ray sources were matched with one or more
potential counterpart. We tried to determine the nature of each
counterpart based on X-ray and two-colour visual photometric data.
Following the method used by Stocke \et (1991) in the Einstein Medium
Sensitivity Survey (EMSS), we constructed a plot of X-ray to visible flux
ratio vs. colour index to try to distinguish extragalactic from galactic
counterparts\footnote{ Stocke \et (1991) advise that the distinction
between galactic coronal and extragalactic sources on such a diagram will
be less for \rosat-band fluxes than for \einstein-band fluxes. Using PIMMS,
we find that the extragalactic-zone boundary needs to be shifted only by
0.13 towards lower \logfx in converting from \einstein\ to \rosat\ bands,
which is smaller than the typical 1$\sigma$ error of
25 per cent in the X-ray flux. Fig.~3 of Motch \et (1998) also indicates clear separation of
coronal and extragalactic sources.}, and then plotted a colour-magnitude
diagram to try to identify possible cluster members (c.f.\ Jeffries
1997).

\begin{figure*}
\begin{center}
\rotatebox{-90}{\includegraphics[width=8cm]{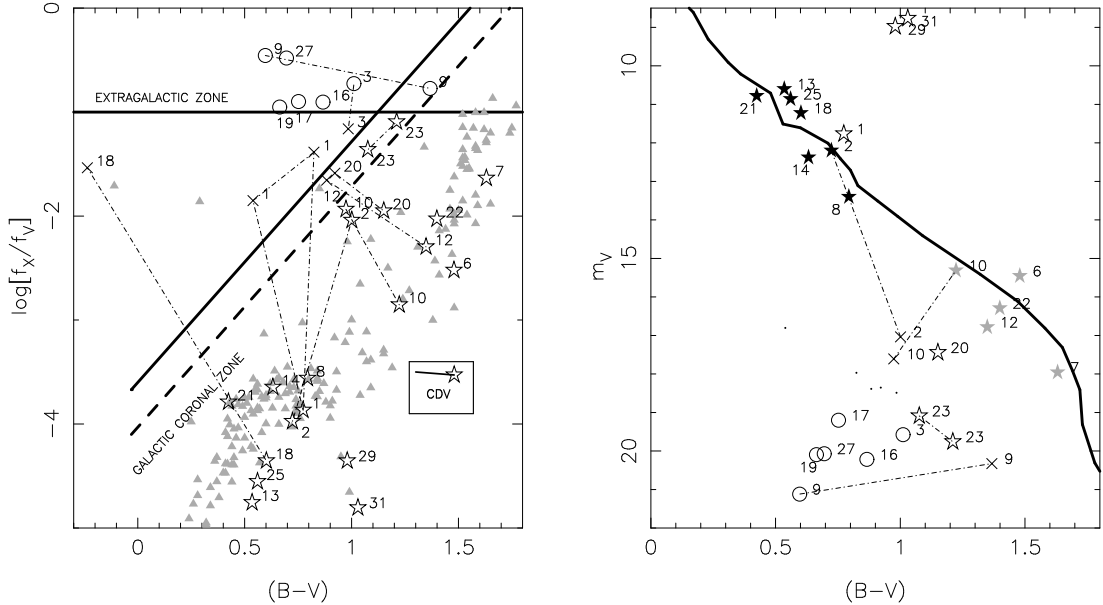}}
\caption{(a) \logfx\ vs \bv and (b) $V$ vs \bv for potential counterparts
to X-ray sources in the field of NGC~6633. The plotted quantities are
not corrected for reddening or absorption. Dot-dashed lines join
candidates that fall within the error circle of the same
source. Numbers refer to the associated source. (a) The thick, diagonal solid
and dashed lines are expected to bound zones of the diagram containing respectively 98 and 90
per cent of coronal sources (derived from Fig.~7 in Stocke \et
1991). The thick horizontal line marks the value above which
extragalactic sources are expected to dominate. We classify potential
counterparts by their position on this diagram. Stars mark objects we
classify as `coronal', circles mark those we classify as
`extragalactic' and crosses mark those we do not consider to be
counterparts to an X-ray source. The shaded triangles mark the
positions of Hyades members, demonstrating that our coronal zone does
not exclude even the most active coronae in the Hyades (the two
objects lying outside the coronal zone are white dwarfs). The box
labelled `CDV' shows the
dereddening vector for objects that have suffered the mean absorption
of the cluster. (b) The thick line is a model MS
corrected for the distance and reddening of the cluster. Unless
previously determined as members or non-members by proper-motion
measurements, we classify
the coronal sources according to their apparent proximity to this
MS. Black stars indicate proper-motion members, grey stars proposed
new members, open stars non-members. Other symbols have same meaning
as in (a). Note that objects rejected in (a) are not plotted in (b).}
\label{fig-6633-keele}
\end{center}
\end{figure*}

Fig.~\ref{fig-6633-keele}a shows X-ray--to--visual flux ratio (listed in column 7 of Table~\ref{tbl-ngc6633-opt}) as a function of colour
index (\bv), for optical candidates to the X-ray sources in the
NGC~6633 field. Only those candidates found in the TJH and ACT
catalogues are plotted. Where there is more than one optical object
within the error circle of an X-ray source, those candidates are
joined by a dot-dashed line. The {\it solid horizontal} line indicates
the flux ratio above which {\it extragalactic} sources are expected to
dominate (cf.\ the EMSS results of Stocke \et (1991)).
The {\it solid diagonal} line indicates the approximate upper boundary
expected for {\it galactic coronal} sources, derived from Stocke \et
(1991), Fig.~7, such that this zone contains $\approx$ 98 per cent of the coronal sources
plotted in that figure. The parallel {\it dashed diagonal} line bounds a zone
that contains $\approx$ 90 per cent of those sources.

We stress that the plotted flux ratios were calculated
assuming no absorption and the colour indices were not corrected
for reddening because absorption data are not available for many of
our X-ray counterparts.

The shift of a source in Fig.~\ref{fig-6633-keele}a from the plotted (reddened)
position to its intrinsic position, its {\it dereddening vector}, is
nearly horizontally leftwards as the flux 
ratio is insensitive to absorption until \nh\ of a few 10$^{21}$
cm$^{-2}$. Thereafter the attenuation of the optical flux dominates and
gives the dereddening vector a slope in the same sense as the coronal zone
boundary, but which remains shallower. Our classification of sources by their
reddened positions may be erroneous for objects that fall just inside
the coronal or extragalactic zone boundary, since their intrinsic
positions could lie outside.

We have considered objects that lie in neither the extragalactic nor the
coronal regions of Fig.~\ref{fig-6633-keele}a to be false
counterparts to X-ray sources. We have plotted the Hyades members
(Stern \et 1995) as
shaded triangles in Fig.~\ref{fig-6633-keele}a, which demonstrates
that even the most active Hyads are not excluded from our coronal zone.
The only Hyads that fall outside are white dwarfs, which we would not
expect to detect through the absorbing column to NGC~6633. While X-ray binaries
and CVs may also fall
in this `No Man's Land' these are rather rare objects (c.f.\ Stocke
\et 1991; Motch \et 1997) and we do
not take further account of them here. This approach is supported by the
fact that all such sources (1, 3 and 18) have other possible
counterparts that fall in either the coronal or extragalactic
zones, and only one source (23) with a candidate just within the
coronal boundary has no alternative deeper in the coronal zone. Any
alternatives to the 3
candidates (to sources 16, 17 and 19) that fall just within the
extragalactic zone must be fainter and hence fall deeper in the
extragalactic zone. A candidate counterpart to source 9 falls in the
overlap region: it may be a highly reddened
extragalactic object, or an active K-type dwarf, or
a mismatch; an alternative counterpart lies deep in the
extragalactic zone. In summary, we classify 17 sources as
`coronal', 5 as `extragalactic' and one has a candidate in each
zone. There are 3 cases where two possible counterparts to a
source both have the same classification.


\begin{figure*}
\begin{center}
\rotatebox{-90}{\includegraphics[width=8cm]{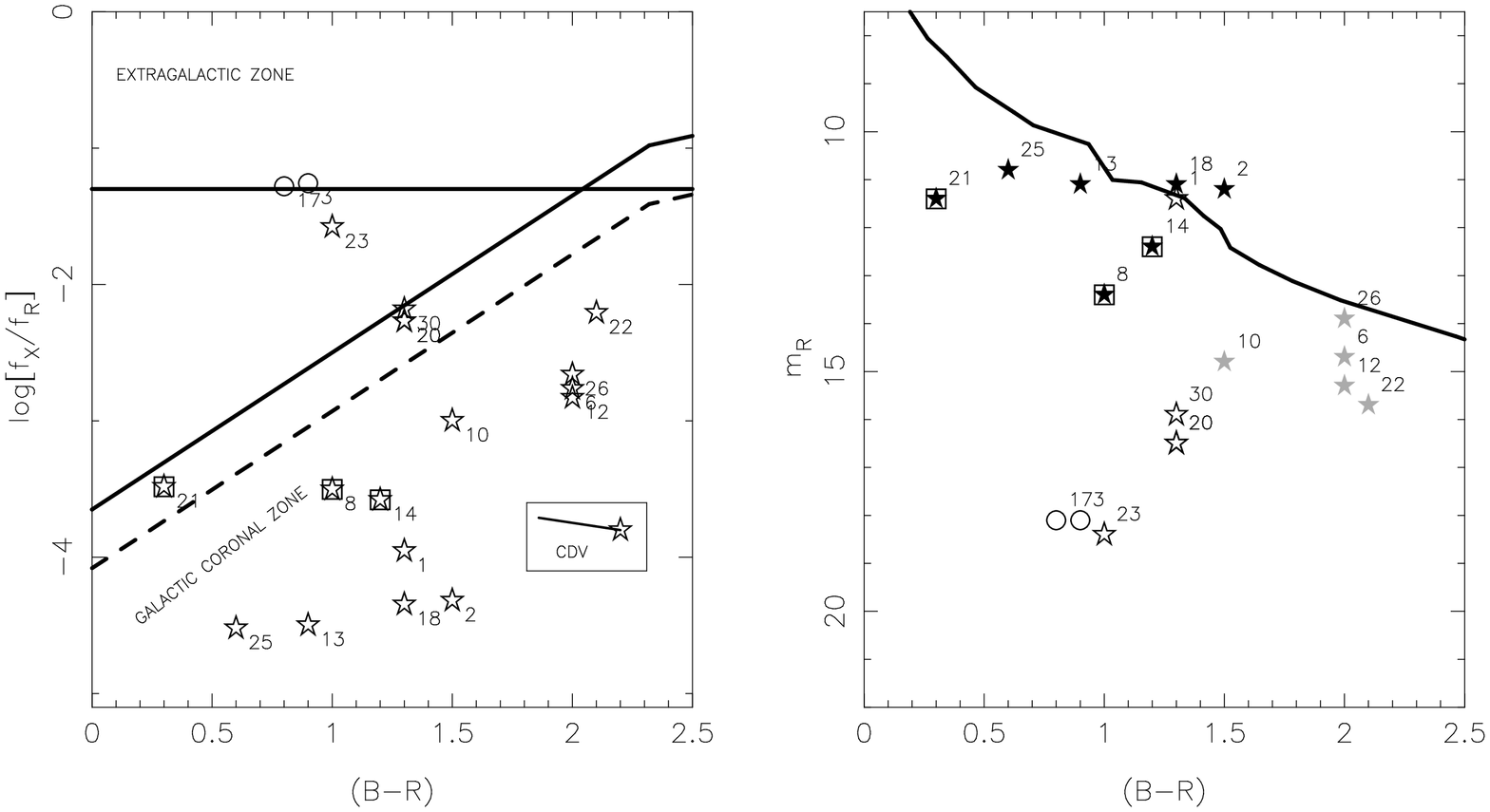}}
\caption{(a) \logfxfr vs \br and (b) $R$ vs \br for potential counterparts
to X-ray sources in the field of NGC~6633. We plot Figures 2a and 2b
for source counterparts found in the USNO-A2.0 Catalogue to establish
the quality of its photometry in classifying potential counterparts. Note
that the giants associated with sources 29 and 31 are not plotted due
to anomalously high colours in this catalogue. (a)
The coronal- and extragalactic-zone boundaries have been converted from
$V$-band to $R$-band as described in the text. Excepting those found
solely in the USNO-A2.0 Catalogue, objects are marked here
according to their classification in Fig.~\ref{fig-6633-keele}a. Stars indicate `coronal'
sources, circles indicate `extragalactic' sources. Squares indicate the
photometry is flagged as `probably wrong' (Monet \et 1998). `CDV' again
indicates the dereddening vector for cluster members.
(b) The empirical MS from
Fig.~\ref{fig-6633-keele}b has been converted from $V$- to $R$-band, as have the applied reddening
and absorption (see text for details). Again, excepting those found
solely in the USNO-A2.0 Catalogue, coronal objects are marked here
according to their classification in Fig.~\ref{fig-6633-keele}b. The symbols have the
same meaning as in 2b, with squares denoting doubtful photometry.}
\label{fig-6633-usno}
\end{center}
\end{figure*}

The counterparts we had not already rejected were plotted on a
colour-magnitude diagram (CMD,
Fig.~\ref{fig-6633-keele}b) to determine which were possible photometric members of the
cluster. Dash-dotted lines connect pairs of counterparts associated
with the same X-ray source. We constructed a model Main Sequence (MS) from an empirical
multicolour survey of MS stars grouped by spectral type (Johnson
1966), and a numerical calculation of absolute magnitude ($M_{\rm V}$) for
each MS spectral type (Schmidt-Kaler 1982). The plotted MS has been 
adjusted for the reddening ($E(B-V)=0.17$) and apparent distance
modulus ($m_{\rm V}-M_{\rm V}=8.0$) of
NGC~6633. The 7 recognised proper-motion members\footnote{We note that only 4 of our `proper motion members' were included
in the membership catalogue used by Harmer \et (2000), as the other 3 fell just outside their strict
criteria: two (possible equal-mass SB2) stars appear 0.2 mag above their
photometric acceptance limit; the third has a membership probability
slightly below their 85 per cent limit.} (counterparts to sources 2, 8, 13, 14,
18, 21, 25) appear within $\sim
0.5$ mag. of the MS
as we would expect. They are
accompanied by the proper-motion non-member (source 1) already mentioned. There
are also 4 fainter counterparts that are $\la$ 1 mag.\ from the MS
(candidates for sources 6, 7, 10 and 22) and 1 further candidate (to
source 12)
within 2 mag.\ of the MS. We noted these as potential photometric
members of NGC~6633. In all cases where one of two possible
counterparts appears close to the MS we took that identification to be
the correct one, as a late-type member of an open cluster is more
likely to be an X-ray emitter than an average (typically older, and in
this case more distant) field star. The two giants (sources 29
and 31) are at the
top of the diagram and there are two coronal sources (20, 23) well
below the MS. 
If the source-9 candidate in the overlap zone of
Fig.~\ref{fig-6633-keele}a is the correct counterpart and is a MS
star, it is $\sim$ 3 kpc distant and has $L_{\rm X} \sim 10^{31}$ \ergs,
which is rather high for a late-type coronal source. It has
$\log(L_{\rm X}/L_{\rm bol}) \sim -2$ (see Table~\ref{tbl-ngc6633-opt}), so
must have been emitting above the saturation limit of
$\log(L_{\rm X}/L_{\rm bol}) \sim -3$, but cannot be ruled out from
being a flaring coronal source. For classification purposes we have
preferred the alternative `extragalactic' counterpart. 


To find counterparts in the 17 per cent of our NGC~6633 field not
covered by the TJH study, we used the USNO-A2.0 catalogue.
The USNO photometry is in $B$ and $R$ bands, derived
from the red and blue POSS--I photographic plates. It is reported to be accurate only
to within 0.2-0.3 magnitudes (Monet \et 1998).  Bright objects (V
$<$ 11) have very poorly determined magnitudes in the USNO Catalogue,
with many flagged as `probably wrong' (denoted by asterisks in column
10 of our Table~\ref{tbl-ngc6633-opt}), but all such objects should have
better photometry available in other catalogues.

19 of the candidates found previously in the TJH/ACT catalogues were
also found in the USNO-A2.0 Catalogue, and there were 2 new
matches: to sources 26 and 30 that were in regions not covered by the
TJH survey.

We have tested whether the USNO photometry is good enough to determine the
nature of source counterparts by plotting \logfxfr\ vs \br and
colour-magnitude diagrams for candidate counterparts to the NGC~6633
sources found in the USNO Catalogue (Fig.~\ref{fig-6633-usno}a, 3b) and comparing the results
to those of the diagrams plotted using \bv CCD photometry (Fig.~\ref{fig-6633-keele}a,
2b). 

The zone boundaries in the \logfxfr\ vs \br diagram were adjusted
for $R$-band photometry.
The coronal-zone boundary was recalculated by fitting the empirical
\bv vs $V-R$ 
values appropriate for MS stars (Johnson 1966) as a broken,
linear relation\footnote{ We use MS relations because the EMSS results
show that most objects near the
coronal-zone boundary are fast-rotating MS stars. 
}.


\begin{table*}
\begin{minipage}{178mm}
\caption{Potential counterparts to X-ray sources in the field of
NGC~6633.}
\label{tbl-ngc6633-opt}
\begin{center}
\scriptsize
\begin{tabular}{rccccccccccrll}
 & \multicolumn{2} {c} {Position} & \multicolumn{4} {l}
{TJH\ddag/ACT data} & \multicolumn{4} {l} {USNO-A2.0 data} & \\
\\
Nr & RA (J2000) & Dec (J2000) & Sep & $m_{\rm V}$ & $B$-$V$ & $[
f_{\rm X}/f_{\rm V}]$ & Sep & $m_{\rm R}$ & $B$-$R$ & $[f_{\rm
X}/f_{\rm R}]$ & $L_{\rm X} \pm \sigma_{L_{\rm X}}$ & Class & Name\\
(1) & (2) & (3) & (4) & (5) & (6) & (7) & (8) & (9) & (10) & (11) & \multicolumn{1}{c}{(12)} & (13) & (14)\\
\\
 1 &18 26 50.92&+ 6 41 49.5&0.85 &11.77 &0.77&-3.87&0.70&11.4&1.3\ &-3.95& 2.5 $\pm$ 1.1&C N PM&S190\\
   &18 26 50.98&+ 6 41 46.2&3.11 &16.81 &0.54&-1.85&    &    &     &     &              &M\\
   &18 26 51.37&+ 6 41 49.0&7.13 &17.97 &0.82&-1.39&    &    &     &     &              &M\\
 2 &18 26 54.75&+ 6 33 30.6&1.18 &12.20 &0.72&-3.98&1.31&11.2&1.5\ &-4.32& 1.3 $\pm$ 0.6&C Y PM&S202\\
   &18 26 54.38&+ 6 33 27.4&5.14 &17.04 &1.00&-2.04&    &    &     &     &              &M\\
 3 &18 26 55.46&+ 6 43 07.0&3.62 &18.49 &0.98&-1.16&3.67&18.1&0.9\ &-1.26&                &E?&0900-12850039\\
   &18 26 55.45&+ 6 43 00.9&4.07 &19.58 &1.01&-0.73&    &    &     &     &                &M\\
 4 &           &           &     &      &    &     &    &    &     &     &                &U\\
 5 &           &           &     &      &    &     &    &    &     &     &                &U\\
 6 &18 27 00.85&+ 6 33 42.6&3.81 &15.46 &1.48&-2.52&4.41&14.7&2.0\ &-2.76& 1.9 $\pm$ 0.6&C Y?Ph&0900-12854908\\
 7 &18 27 03.23&+ 6 39 50.2&2.03 &17.96 &1.63&-1.64&    &    &     &     & 1.4 $\pm$ 0.6&C Y Ph\\
 8 &18 27 07.54&+ 6 27 55.8&3.91 &13.40 &0.79&-3.56&3.78&13.4&1.0* &-3.50& 1.1 $\pm$ 0.5&C Y PM&J27\\
 9 &18 27 08.00&+ 6 35 16.9&5.12 &21.12 &0.60&-0.46&    &    &     &     &                &E\\
   &18 27 07.78&+ 6 35 20.6&1.51 &20.33 &1.37&-0.77&    &    &     &     &                &C N Ph\\
10 &18 27 11.62&+ 6 31 06.5&2.29 &15.31 &1.22&-2.85&2.48&14.8&1.5\ &-3.00& 1.0 $\pm$ 0.5&C Y Ph&0900-12863797\\
   &18 27 11.44&+ 6 31 07.3&3.68 &17.61 &0.97&-1.94&    &    &     &     &                &M\\
11 &           &           &     &      &    &     &    &    &     &     &                &U\\
12 &18 27 18.90&+ 6 35 39.5&4.00 &16.79 &1.35&-2.30&3.76&15.3&2.0\ &-2.83& 0.9 $\pm$ 0.5&C Y?Ph&0900-12870017\\
   &18 27 18.85&+ 6 35 40.7&3.29 &18.39 &0.88&-1.66&    &    &     &     &                &M\\
13 &18 27 20.19&+ 6 38 54.5&0.54 &10.60 &0.54&-4.76&0.82&11.1&0.9\ &-4.50& 0.9 $\pm$ 0.5&C Y PM&S279\\
14 &18 27 26.40&+ 6 34 18.3&2.56 &12.38 &0.63&-3.65&2.10&12.4&1.2* &-3.58& 2.4 $\pm$ 0.6&C Y PM&J34\\
15 &           &           &     &      &    &     &    &    &     &     &                &U\\
16 &18 27 35.57&+ 6 38 11.9&1.91 &20.22 &0.87&-0.90&    &    &     &     &                &E?\\
17 &18 27 39.47&+ 6 25 58.2&1.38 &19.20 &0.75&-0.90&1.77&18.1&0.8\ &-1.28&                &E?&0900-12886981\\
18 &18 27 43.43&+ 6 41 11.4&3.57 &11.23 &0.60&-4.35&3.37&11.1&1.3\ &-4.35& 1.3 $\pm$ 0.6&C Y PM&J25 SB\\\
   &18 27 43.63&+ 6 41 13.1&4.36 &18.27 &0.24&-1.54&    &    &     &     &                &M\\
19 &18 27 44.31&+ 6 38 50.3&2.93 &20.09 &0.66&-0.95&    &    &     &     &                &E?\\
20 &18 27 44.87&+ 6 37 20.5&3.24 &17.44 &1.15&-1.95&3.33&16.5&1.3\ &-2.27&$>$1.1 $\pm$ 0.5&C N Ph&0900-12891233\\
   &18 27 44.84&+ 6 37 23.7&4.41 &18.35 &0.92&-1.59&4.80&16.5&1.3\ &-2.27&                &M&0900-12891208\\
21 &18 27 45.49&+ 6 29 23.6&1.62 &10.78 &0.42&-3.79&1.77&11.4&0.3* &-3.48& 7.4 $\pm$ 1.0&C Y PM&S359 SB\\
22 &18 27 46.53&+ 6 45 39.8&2.74 &16.30 &1.40&-2.03&3.01&15.7&2.1\ &-2.21& 2.7 $\pm$ 1.0&C Y Ph&0900-12892577\\
23 &18 27 51.97&+ 6 35 10.7&2.52 &19.09 &1.08&-1.36&2.67&18.4&1.0\ &-1.58&$>$0.9 $\pm$ 0.5&C?N Ph&0900-12896782\\
   &18 27 51.91&+ 6 35 11.5&3.42 &19.76 &1.21&-1.09&    &    &     &     &$>$0.9 $\pm$ 0.5&C?N Ph\\
24 &           &           &     &      &    &     &    &    &     &     &                &U\\
25 &18 28 00.46&+ 6 31 45.2&1.24 &10.86 &0.56&-4.55&1.87&10.8&0.6\ &-4.52& 1.2 $\pm$ 0.5&C Y PM&S407\\
26 &18 28 00.61&+ 6 20 59.3&     &      &    &     &5.15&13.9&2.0\ &-2.66& 4.9 $\pm$ 1.6&C Y?Ph&0900-12903693\\
27 &18 28 01.78&+ 6 25 46.1&1.27 &20.07 &0.70&-0.48&    &    &     &     &                &E\\
28 &\multicolumn{8}{l}{Spurious detection (see text for details)}\\
29 &18 28 17.64&+ 6 46 00.0&6.19 & 8.98 &0.98&-4.35&6.20& 8.3&4.0* &-4.57&10.6 $\pm$ 1.9&C N?PM&BD+06 3796\\
30 &18 28 19.27&+ 6 35 12.3&     &      &    &     &5.13&15.9&1.3\ &-2.18&$>$2.4 $\pm$ 0.9&C?N Ph&0900-12919311\\
31 &18 28 22.99&+ 6 42 29.1&4.39 & 8.77 &1.03&-4.81&4.61& 8.0&3.3* &-5.05& 4.6 $\pm$ 1.6&C N?PM&BD+06 3798\\
32 &           &           &     &      &    &     &    &    &     &     &                &U\\
\end{tabular}
\vspace{-0.5cm}
\footnotetext{Key:
(1) X-ray source number
(2 and 3) Optical source position (J2000) from USNO-A2.0 where
available, else from TJH;
(4 and 8) Separation from source position (arcsec);
(7) $\log(f_{\rm X}/f_{\rm V})$; 
(11) $\log(f_{\rm X}/f_{\rm R})$;
(12) 0.1-2.4 keV X-ray luminosity and uncertainty (10$^{29}$ erg/s).
(13) Classification. Type: C=`coronal', E=`extragalactic',
M=considered to be incorrect identification, U=unidentified; Membership: Y/N; Method: PM=proper
motion, Ph=photometric; `?' denotes substantial doubt. The proposed new
members are indicated by `C Y Ph' or `C Y?Ph'.
(14) Name: numbering from Jeffries (1997; J), Sanders (1973; S) and USNO-A2.0(0900-);
SB=spectroscopic binary. \ddag TJH refers to the survey carried out by
Totten, Jeffries \& Hambly (2000).}
\end{center}
\end{minipage}
\end{table*}


To recalculate the `extragalactic-zone' boundary, we used a
photometric survey of 27 bright ($V \la 17$) radio galaxies (Machalski
\& Wi\'{s}niewski 1988) to find a mean extinction-corrected $V-R$ of
0.9, corresponding to a shift of $-0.3$ from \logfx\ to \logfxfr.

When we plotted \logfxfr\ (listed in column~11 of Table~\ref{tbl-ngc6633-opt}) vs \br (Fig.~\ref{fig-6633-usno}a),
for potential counterparts found in USNO-A2.0,
all matches common with the TJH/ACT candidates were found to have the
same classification, excepting the source 23 candidate, which appeared
in the `No-Man's Land' region of the diagram, and not borderline
coronal. Counterparts to sources 20 and 21, although still within the
coronal zone, appear much
closer to the coronal boundary than in Fig.~\ref{fig-6633-keele}a. The photometry of the source 21
counterpart (as those
of sources 8 and 14) is flagged as `probably wrong'. We conclude that the
USNO photometry is sufficient to determine the general classification
of the source.

\begin{figure*}
\begin{center}
\rotatebox{-90}{\includegraphics[width=8cm]{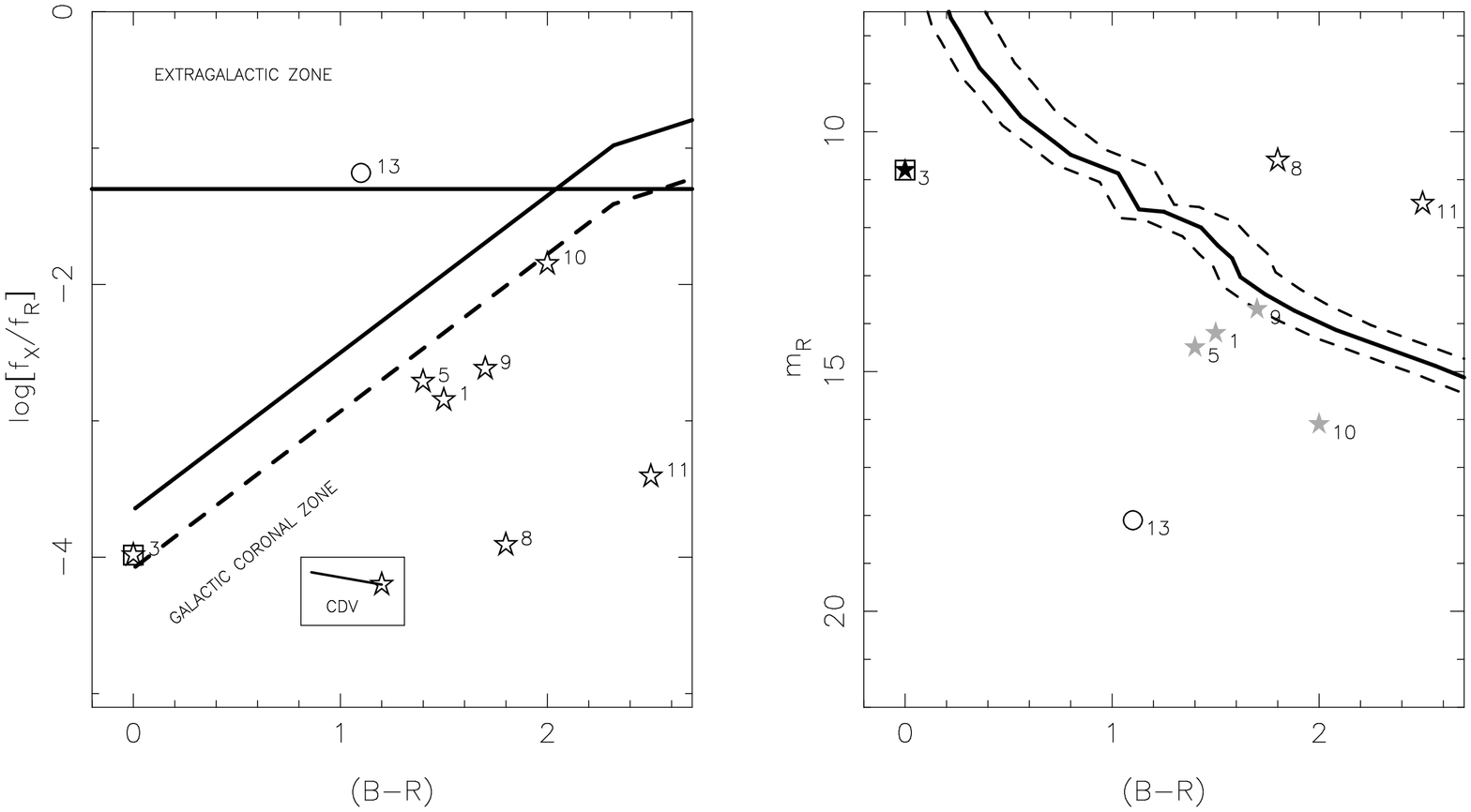}}
\caption{(a) \logfxfr vs \br and (b) $R$ vs \br for potential counterparts
to X-ray sources in the field of IC~4756. We attempt to classify
these counterparts using their USNO-A2.0 photometry as in
Figs.~3a,~b. Numbers refer to the associated source. 
(a) The zone boundaries and symbols are the same as in
Fig.~\ref{fig-6633-usno}a, except classification is via this diagram alone. Once more,
the cluster dereddening vector (CDV) is indicated.
(b) The model MS is the same one used in Fig.~\ref{fig-6633-usno}b and a
consistent distance modulus of 8.0 has been applied, but the variable 
reddening across the cluster field is
accounted for. The solid line represents the mean value of
$E(B-V)=0.20$ and the dashed lines mark the extremes of the variation
(Schmidt 1978). The black star marks the single proper motion member,
the open stars indicate non-members. Grey stars indicate those sources that
appear in a part of the diagram in which proposed NGC~6633 members
fell in Fig.~\ref{fig-6633-usno}b. Other symbols have the same meaning as in Fig.~\ref{fig-6633-usno}b.}
\label{fig-4756-usno}
\end{center}
\end{figure*}

We plotted the counterparts found in the USNO database on an $R$
vs. \br colour-magnitude diagram (Fig.~\ref{fig-6633-usno}b). Note that the two
giants are not plotted as their USNO \br values are $>$ 3.0 and
`probably wrong'. A model MS in this band was constructed
in a similar way to
the Fig.~\ref{fig-6633-keele}b MS from numerical $M_{\rm V}$ calculations
(Schmidt-Kaler 1982) and empirical \bv, $V-R$ colour values
(Johnson 1966). The $V-R$ values were tested by superposing the model MS in
these colours onto a plot of Praesepe members down to $V \sim 17$ on a $V$ vs $V-R$
diagram (data from Upgren, Weis \& de Luca 1979; Weis 1981; Stauffer 1982 and
Mermilliod \et 1990, converted from Kron to Johnson $V-R$ using colour
conversions of Bessell (1987) and Fernie (1983)). A good observational
fit was found. An intrinsic distance modulus ($m_{\rm V_{0}}-M_{\rm
V_{0}}$) of 7.5 and absorption
$A_{\rm R} = 0.7 A_{\rm V}$ were taken to set the MS at the expected
position for NGC~6633. Note the proper-motion members are
scattered much more widely around the MS; even those whose photometry
is not flagged as `probably wrong' diverge by as much as 1.5 mag.\ (or
0.4 mag.\ in colour). The fainter coronal counterparts that
appeared as `possible photometric members' from their TJH magnitudes now
mainly appear well below the MS (sources 6, 10, 12 and 22). We do not
know the cause of this apparently systematic effect.
We conclude that the
USNO photometry is not good enough to identify sources as probable
photometric members of the cluster, although the candidate
counterpart to source 26
appears close enough to the MS to be interesting, and it seems sufficient
to denote some stars as confident non-members -- as in the case of the
source 30 candidate counterpart.  Our proposed
classification for each source appears in column 13 of Table~\ref{tbl-ngc6633-opt}, and is
based on the ACT/TJH photometry where possible. We consider these
classifications to be sound for $V \la 17$ (82 per cent of the
background sources are fainter than this). In a region so close to
the Galactic plane, there is a strong
probability of fainter objects falling into an X-ray error circle by
chance. We expect to be able to identify even fainter potential cluster
members ($B-V \sim 1.7$), as most (85 per cent) of the
background objects have $B-V < 1.2$. We have
calculated 0.1--2.4 keV luminosities for `coronal' sources, using an
intrinsic count-to-flux conversion factor (\ctofint) of 5.3$\times 10^{-11}$ erg cm$^{-2}$ s$^{-1}$ per HRI count/s,
and assuming a cluster distance of 320 pc. This \ctofint\ factor was
calculated for a 1.4 keV Raymond-Smith plasma, with $N_{\rm
H}=1\times 10^{21}$ cm$^{-2}$ using PIMMS. A 0.86 keV plasma with the
same column gives a \ctofint\ of 4.5$\times 10^{-11}$ erg
cm$^{-2}$ s$^{-1}$ per HRI count/s. We list these luminosities in
column 12 of Table~\ref{tbl-ngc6633-opt}, indicating upper and lower limits for
foreground and background objects respectively, as appropriate. We
also list errors in these values, calculated from
individual count-rate errors and allowing for a 25 per cent
uncertainty in \ctofint\ and a 10 per cent uncertainty in the cluster
distance. The Hipparcos parallax of NGC~6633 (Robichon \et
1999) is based on only 4 members, and has too large an uncertainty to
calculate a useful distance.

\subsection{IC~4756}

\begin{table*}
\begin{minipage}{178mm}
\caption{Potential counterparts to X-ray sources in the field of
IC~4756.}
\label{tbl-ic4756-opt}
\begin{center}
\scriptsize
\begin{tabular}{rccccccccccrll}
 & & & \multicolumn{4} {l}
{ACT data} & \multicolumn{4} {l} {USNO-A2.0 data} & \\
\\
Nr & RA (J2000) & Dec (J2000) & Sep & $m_{\rm V}$ & $B$-$V$ & $[{
f_{\rm X}}/{f_{\rm V}}]$ & Sep & $m_{\rm R}$ & $B$-$R$ & $[{f_{\rm X}}/{f_{\rm R}}]$ & $L_{\rm X} \pm \sigma_{L_{\rm X}}$ & Class & Name\\
(1) & (2) & (3) &  (4) & (5) & (6) & (7) & (8) & (9) & (10) & (11) & \multicolumn{1}{c}{
(12)} & (13) & (14)\\
\\
 1 &18 38 20.36 & +05 25 00.9&      &       &      &       & 2.52 & 14.2  & 1.5& -2.85 & 3.8 $\pm$ 1.2 &C Y?Ph& 0900-13330140\\
 2 &		&	     &      &       &      &       &      &       &    &       &           &U \\
 3 &18 38 35.87 & +05 34 49.4& 3.23 &  9.90 & 0.41 & -4.41 & 2.57 & 10.8* & 0.0& -3.98 & 6.3 $\pm$ 1.6 &C Y PM & BD~+05~3863\\
 4 &		&	     &      &       &      &       &      &       &    &       &           &U \\
 5 &18 38 38.04 & +05 33 56.5&      &       &      &       & 2.63 & 14.5  & 1.4& -2.71 & 3.9 $\pm$ 1.2 &C Y?Ph& 0900-13341340\\
 6 &		&	     &      &       &      &       &      &       &    &       &           &U \\
 7 &		&	     &      &       &      &       &      &       &    &       &           &U \\
 8 &18 38 51.16 & +05 31 52.0& 2.83 & 10.95 & 0.33 & -3.83 & 2.49 & 10.6  & 1.8& -3.91 &$<$ 9.0 $\pm$ 1.8 &C N PM & HSS~247\\
 9 &18 39 04.96 & +05 19 35.0&      &       &      &       & 0.20 & 13.7  & 1.7& -2.62 &10.2  $\pm$ 2.4 &C Y?Ph& 0900-13359235\\
10 &18 39 05.93 & +05 35 01.8&      &       &      &       & 2.84 & 16.1  & 2.0& -1.85 & 6.6 $\pm$ 1.8 &C Y?Ph& 0900-13359846\\
11 &18 39 05.90 & +05 34 26.0&      &       &      &       & 2.51 & 11.5  & 2.5& -3.41 &$<$12.6 $\pm$ 2.2 &C N Ph & GSC~0455~00727\\
12 &		&	     &      &       &      &       &      &       &    &       &           &U \\
13 &18 39 22.03 & +05 22 52.0&      &       &      &       & 4.46 & 18.1  & 1.1& -1.18 &           &E? & 0900-13370352\\
\end{tabular}
\footnotetext{Key:
(1) X-ray source number;
(2 and 3) Optical source position (J2000) from USNO-A2.0;
(4 and 8) Separation from source position (arcsec); 
(7) $\log({f_{\rm X}}/{f_{\rm V}})$;
(11) $\log({f_{\rm X}}/{f_{\rm R}})$;
(12) 0.1--2.4 keV luminosity and 1$\sigma$ uncertainty (10$^{29}$ erg/s);
(13) Classification (as for Table~\ref{tbl-ngc6633-opt});
(14) Name: USNO-A2.0(0900-).}
\end{center}
\end{minipage}
\end{table*}

Of their 10 X-ray sources, Randich \et (1998) could identify three with
optical counterparts: one (BD~+05~3863) a member of the cluster; the
second (HSS~247) a galactic non-member; the third a $V\sim15$ object listed
in the GSC. We have attempted, employing the
method described above with the USNO-A2.0 Catalogue, to improve the
identification of sources in the IC~4756 field.

The USNO-A2.0 Catalogue was searched for optical matches within 12 arcsec
of the X-ray source positions; this indicated a systematic shift (of +4.3
and $-$0.5 arcsecs in RA and Dec respectively) that was added to each X-ray
source position. The USNO-A2.0 Catalogue was then searched again for
optical matches within the error circles of the X-ray sources. 8 were
found and their positional and photometric data appear in columns
2--3 and 8--10 of Table~\ref{tbl-ic4756-opt} (there
were no multiple-candidate cases). When plotted on a \logfxfr\ vs \br
diagram (Fig.~\ref{fig-4756-usno}a), 7 are suggested as
coronal sources, and one (source 13) as `extragalactic'. When plotted on a
CMD (Fig.~\ref{fig-4756-usno}b), the coronal sources are scattered
widely around an empirical MS. The
positions of the counterparts to sources 1, 5 and 9 on the diagrams
are similar to that of the NGC~6633 source 10, and so cannot be ruled
out as potential cluster
members; nor can source 10, which occupies a similar position to
NGC~6633 source 22 in the CMD. BD~+05~3863 --- the cluster member identified as an X-ray source
in R98 (and corresponding to our source 3) --- sits well below the
MS but its magnitudes are flagged as `probably wrong' in
USNO-A2.0. This star and HSS~247 --- the non-member identified by R98,
corresponding to our source 8 --- are included in the Tycho (ACT)
Catalogue (see columns 4--7 of Table~\ref{tbl-ic4756-opt} for data). Their positions on a \logfx\ vs \bv diagram confirm their
coronal nature, and both appear within 0.5 mag.\ of the mean MS on a V vs \bv
plot. HSS~247 has a \br value much larger than that expected for a
main-sequence star with its Tycho photometry, despite its USNO magnitudes
being unflagged. It appears that the USNO photometry of any bright star
($V \la 11$) should be treated with caution. GSC~0455~00727 --- the
third identification made by R98, counterpart to our source 11 --- appears well above the IC~4756
MS, indicating that it is a foreground object. Our proposed
classification for each source appears in column 13 of Table~\ref{tbl-ic4756-opt}. We consider our
proposed classifications to be
very tentative. We
calculate 0.1--2.4 keV luminosities for coronal sources, using a \ctofint\ of 5.3$\times 10^{-11}$ erg cm$^{-2}$ s$^{-1}$,
and assuming a cluster distance of 400 pc, indicating upper limits for
foreground objects as appropriate, and list these and associated
errors (estimated as for NGC~6633) in column 12 of
Table~\ref{tbl-ic4756-opt}. The recently derived Hipparcos distance of 330 pc, based on
the parallaxes of 9 members, has an
associated error of 20 per cent (Robichon \et 1999).



\section{Comparisons with simulations of the Hyades}




We have compared NGC~6633 and IC~4756 to the Hyades by simulating
observations of the Hyades at the distance and sky position of each
cluster. The X-ray luminosity of each cluster member was taken from the Hyades \rosat\
All-Sky Survey (RASS)
Catalogue (Stern, Schmitt \& Kahabka 1995): undetected early-type stars were
assigned zero luminosities, while undetected K and M dwarfs
($B-V~>$~0.8) were assigned luminosities at random from the Hyades
pointed-survey luminosity functions (Pye \et 1994).


We also simulated the field sources expected at the
position in the sky of each cluster. We assumed the field sources would
be comprised of galactic coronal sources (late-type stars) and
extragalactic sources (mainly active galaxies). Galactic sources were
simulated using the flux distribution
(\logns) constructed by Guillout (1996), appropriate for
the sky position of the simulated field. 475 sources per square degree
brighter than $\rm 10^{-15}$ erg cm$^{-2}$ s$^{-1}$ were simulated, and
the X-ray flux of each was selected at random from the \logns.
Extragalactic sources were simulated using the broken power-law flux
distribution (\logns) of Hasinger \et (1993), attenuated by a
galactic transmission factor (GTF $\le$ 1.0) to account for absorption
along the line of sight through the Galaxy. For each cluster, we used
the $N_{\rm H}$ due to atomic hydrogen (Dickey \& Lockman 1990) and
added the mean
$N_{\rm H}$ due to molecular hydrogen (Dame \et 1987) in that
direction, getting 3$\times 10^{21}$ cm$^{-2}$ for NGC~6633 and
5$\times 10^{21}$ cm$^{-2}$ for IC~4756. 685 sources per square degree
brighter than $10^{-15}$ erg cm$^{-2}$ s$^{-1}$ were simulated, and the
X-ray flux of each was selected at random from the \logns. The
position of each field source was computed from a uniform random
distribution.


We used PSS to construct `sensitivity' maps from the NGC~6633 and
IC~4756 \rosat\ HRI images and backgrounds. 
The sensitivity map shows the minimum detectable point-source count rate
at each location in the HRI image, for the selected significance
threshold (as summarised in Table~\ref{tbl-sen}).
The flux of each simulated source was
compared to the value of the sensitivity map at the source
position. Sources within 17 arcmin of the field centre whose flux exceeded the minimum detectable flux at
their position were flagged as detections. The numbers of detected
sources of each source type were compared to the samples we obtained
from the HRI observations. The results are summarised in
Table~\ref{tbl-sim}. The range in simulated field-source detections is the
$1\sigma$ confidence interval of multiple simulations, while the range
in simulated cluster-source detections simply applies a $\sqrt{N}$ error to the
fixed number of simulated cluster detections. Note that this is a
`perfect' simulation as we know {\it a priori} the type of each
simulated source. We have not simulated the optical properties of the
field sources, and so do not make predictions on possible errors made in
classification by the method used in \S3; all cluster
members fall well within the coronal zone.

\begin{table}
\begin{minipage}{85mm}
\caption{Comparison of numbers of X-ray sources: ($N_{\rm sim}$) predicted as
detectable by our
simulations of the Hyades at these cluster distances 
with ($N_{\rm class}$) those detected in the \rosat\ HRI images and classified using
optical two-colour photometry.}
\label{tbl-sim}
\begin{center}
\scriptsize
\begin{tabular}{llcccc}
 & &\multicolumn{2}{c}{\bf NGC~6633}&\multicolumn{2}{c}{\bf IC~4756}\\
\multicolumn{2}{l}{Distance (pc)} &\multicolumn{2}{c}{320}&\multicolumn{2}{c}{400}\\
\multicolumn{2}{l}{Exposure time (ks)} &\multicolumn{2}{c}{119}&\multicolumn{2}{c}{88}\\
\multicolumn{2}{l}{Source Type} & $N_{\rm sim}$ & $N_{\rm class}$ & $N_{\rm sim}$ & $N_{\rm class}$\\
  &  \\
\\
\multicolumn{2}{l}{\bf `Coronal' sources} & 18--31 & 19 & 8--18 & 7\\
 & Cluster sources & 10--17 & 13 & 5--11 & 5\\
 & Galactic field sources  & 8--14 & 6 & 4--9 & 2\\
\multicolumn{2}{l}{\bf `Extragalactic' sources} & 2--7 & 6 & 0--5 & 1\\
\multicolumn{2}{l}{\bf Unidentified sources} &  & 6 & & 5\\
\\
\multicolumn{2}{l}{\bf Total number of sources} & 20--40 & 31 & 8--23 & 13\\
\end{tabular}
\end{center}
\end{minipage}
\end{table}


In agreement with our simulations (Table~\ref{tbl-sim}), cluster members are the
dominant source type in both measured fields. At first sight, the
agreement between the predicted and classified numbers of cluster
members is good, and suggests that NGC~6633 in particular has
X-ray properties similar to our template cluster the Hyades. However,
our simulation takes no account of differences in {\it richness}
between clusters, i.e.\ the spatial density of cluster
members. The incompleteness of membership lists means we can
compare the richness of NGC~6633 and the Hyades only for F- and G-type
stars. Harmer \et (2000) find 15 members in the range
0.4$\leq (B-V)_0 < 0.8$ within 17 arcmin of the
field centre. When we account for the 83 per cent coverage of the TJH study,
this figure is increased to 18. Note that Harmer \et used strict proper motion and
photometric criteria for membership, and so near-equal mass binaries
in particular will have been excluded from this number. We find 19
members in this colour range in the field of the Hyades at 320 pc (which includes binaries (Stern \et 1995)) suggesting the two
clusters are of similar richness (at least in F- and G-type members,
although there is a $\sim$ 30 per cent uncertainty in the counting
statistics). The agreement in detected members between our HRI study
of NGC~6633 and our simulation of NGC~6633 as the Hyades at 320 pc
suggests that the proportion of members that emit X-rays is
similar in both clusters. This analysis does not take into account the
X-ray luminosities of these detected members. Harmer \et (2000) and
Franciosini \et (2000) both find the mean level of X-ray activity
in NGC~6633 to be lower than that of the Hyades. None of the sources we classify as a 
member of NGC~6633 has an X-ray luminosity as high as 10$^{30}$ erg
s$^{-1}$, so the cluster appears to be lacking a population of
highly active stars like those seen in the Hyades and Praesepe. 

Membership lists for IC~4756 are less complete than those for
NGC~6633, but the list given in R98 suggests a richness
in the aforementioned F-G-type colour range similar if not slightly
lower than the Hyades (13 MS members). Given the
uncertainties in the USNO-A2.0 photometry, it is difficult to make a
judgement on the number of detected members in IC~4756, but the
uncertain nature of 4 of our 5 classified `members' may point more
toward a lower than a higher activity level than that of the Hyades.

We cannot draw comparisons between the simulated and classified numbers
of field sources to meaningfully constrain the number-flux
distributions employed. The ratio of classified extragalactic to
coronal sources is higher than expected from simulations, but
the high surface density
of faint stars this close to the plane (b = 8.3$^{\circ}$) suggests that
at least half of our candidate `extragalactic' associations are
erroneous, reducing such sources to the status of the 5 unidentified
ones. The completeness level of the TJH survey is $V\sim 19$, so these
unidentified
objects must have \logfx\ $> -1.5$. They may still be extragalactic, or
faint, red, fast-rotating coronal sources --- dK or dM stars, some of
which {\it could} be cluster members.

\section{The Potential of XMM for studying Hyades-age Open Clusters}

The problem of differing X-ray activity among coeval open clusters
remains. Two hindrances are apparent: this work suggests that faint,
as yet unrecognised members contribute considerably to X-ray emission
from Hyades-age clusters, so the incompleteness of membership lists is
the first problem. Ongoing spectroscopic and photometric work by
Jeffries \et (e.g.\ Totten, Jeffries \& Hambly 2000) may
resolve this problem for NGC~6633, and a study by Upgren \et (Upgren,
Lee \& Weis 1998) should
do likewise for IC~4756. Secondly, X-ray studies of these two clusters
(this work; Harmer \et 2000; R98) have
demonstrated that the obtained \rosat\ observations were not sensitive enough to study intermediate-age open
clusters beyond the nearby Hyades, Praesepe and Coma systems down to the
emission level at which comprehensive comparisons could be made. We do
not know if the Hyades or Praesepe should be considered the archetype,
or whether a continous spread of activity levels means there is no archetype.

We have produced a simulation to test if the recently launched \xmm\
X-ray observatory is capable of probing NGC~6633 and IC~4756 down to a
sufficiently deep activity level in a reasonable exposure time. We constructed a sensitivity map appropriate for a
5$\sigma$ detection threshold in a 50~ks \xmm\
EPIC pn observation, by assuming a background rate of 2.0$\times10^{-6}$
count s$^{-1}$ arcsec$^{-2}$ and using a sliding detection cell with diameter
equal to the nominal PSF HEW of 15 arcsec (vignetting is also taken into
account). As in \S 4 with the HRI sensitivity map, our
simulation of the X-ray characteristics of the Hyades at
300~pc was compared to this map to determine the detection rate of
members of different spectral types, and the number of
contaminating field sources detected. We find 44 cluster, 47 galactic
field and 26 extragalactic field sources were detected. All 15 F-
and G-type stars in the field were detected, 71 per cent of K-dwarfs and 61 per cent of
M-dwarfs, indicating that such an \xmm\ observation could reach a
sensitivity comparable to that achieved by \rosat\ with the Hyades. 

\section{Summary}

We have studied deep \rosat\ HRI observations of the Hyades-age open
clusters NGC~6633 and IC~4756, primarily to search for 
unrecognised cluster members.
Based on X-ray (photometric) data and high-quality (CCD) optical two-band
photometry alone, we have classified
the majority of the 31 sources detected in the field of NGC~6633,
proposing 6 possible new members and demonstrating that field sources
do not strongly confuse such efforts, down to $m_{\rm V}$ $\sim$18. 

An attempt to classify these same sources using photometry listed in
the USNO-A2.0 deep, all-sky catalogue has indicated that this photometry is
sufficient to distinguish coronal from extragalactic sources, but
not to confidently establish photometric membership of the cluster.

We have used USNO-A2.0 to propose optical counterparts for 8 of the 13
X-ray sources detected in the field of IC~4756, and indicated 4 objects
whose available photometry appears consistent with cluster membership.

A novel comparison with the Hyades has been made by predicting the
number of sources the \rosat\ HRI would detect if it observed the
Hyades, set at the distances and galactic latitudes of NGC~6633 and
IC~4756, for the respective exposure times granted those two clusters. We have
found good agreement between these predicted numbers and our classified
numbers of sources, with cluster members composing $\sim$ 40 per cent
in each case. This
result suggests that NGC~6633 and IC~4756 are consistent with having a similar
proportion of X-ray emitting stars to the
Hyades, although there is considerable uncertainty in the relative
richness of these clusters, particularly for K- and M-type stars. Both clusters appear to lack the populations of highly
active F- and G-stars seen in the Hyades and Praesepe.

Our simulation of an NGC~6633-distance Hyades predicts that
a 50 ks \xmm\ observation would reach activity levels in NGC~6633
similar to those probed by \rosat\ in the Hyades, enabling a comprehensive
comparison to be made between these intermediate-age clusters.

\section*{Acknowledgments}
KRB and JPP acknowledge the financial support of the UK Particle Physics
and Astronomy Research Council. This work made use of archival material
from the SIMBAD and VIZIER systems at CDS, Strasbourg, and the
Leicester Database and Archive Service (LEDAS). The Digitized Sky
Survey was produced at the Space Telescope Science Institute, under US
Government grant NAG W-2166 from the original National
Geographic--Palomar Sky Survey Plates.

{}

\label{lastpage}

\end{document}